\documentclass[aps,showpacs,showkeys]{revtex4}

\usepackage{amssymb,amsmath}
\usepackage[dvips]{graphicx}

\begin{document}

\title{Phase synchronization between collective rhythms of fully locked oscillator groups}

\author{Yoji Kawamura}
\email{ykawamura@jamstec.go.jp}
\affiliation{Department of Mathematical Science and Advanced Technology,
Japan Agency for Marine-Earth Science and Technology, Yokohama 236-0001, Japan}

\date{April 30, 2014}     



\begin{abstract}
  We study the phase synchronization between collective rhythms of fully locked oscillator groups.
  For weakly interacting groups of two oscillators with global sinusoidal coupling,
  we analytically derive the collective phase coupling function,
  which determines the dynamics of the collective phase difference between the groups.
  We demonstrate that the groups can exhibit anti-phase collective synchronization
  in spite of microscopic in-phase external coupling and vice versa.
\end{abstract}

\maketitle

\begin{quotation}
{\bf
  A system of coupled oscillators can exhibit a rich variety of dynamical behaviors.
  When we investigate the dynamical properties of the system,
  we first analyze individual oscillators and the microscopic interactions between them.
  However, the structure of a coupled oscillator system is often hierarchical,
  so that the collective behaviors of the system cannot be fully clarified
  by simply analyzing each element of the system.
  For example, we found that two weakly interacting groups of coupled oscillators
  can exhibit anti-phase collective synchronization between the groups
  even though all microscopic interactions are in-phase coupling.
  This counter-intuitive phenomenon can occur
  even when the number of oscillators belonging to each group is only two,
  that is, when the total number of oscillators is only four.
  In this paper, we clarify the mechanism underlying this counter-intuitive phenomenon
  for two weakly interacting groups of two oscillators with global sinusoidal coupling.
}
\end{quotation}

\section{Introduction} \label{sec:1}


A system of coupled oscillators provides abundant examples of dynamical behaviors including synchronization phenomena
\cite{ref:winfree80,ref:kuramoto84,ref:pikovsky01,ref:strogatz03,ref:manrubia04,ref:osipov07,ref:mikhailov13,
  ref:hoppensteadt97,ref:izhikevich07,ref:ermentrout10,ref:schultheiss12}.
Among them,
collective synchronization emerging from coupled phase oscillators has been widely investigated
not only for globally coupled systems but also for complex network systems
\cite{ref:strogatz00,ref:acebron05,ref:boccaletti06,ref:arenas08,ref:dorogovtsev08,ref:barrat08}.
Furthermore,
the dynamical behaviors exhibited by interacting groups of globally coupled phase oscillators
have been intensively investigated~\cite{ref:okuda91,
  ref:montbrio04,ref:abrams08,ref:barreto08,ref:sheeba08,ref:sheeba09,ref:laing09,ref:skardal12,ref:anderson12,ref:laing12}.
The appearance of the Ott-Antonsen ansatz~\cite{ref:ott08,ref:ott09,ref:ott11}
has considerably facilitated theoretical investigations
on interacting groups of noiseless nonidentical phase oscillators with global sinusoidal coupling.
In addition,
interacting groups of globally coupled phase oscillators as well as a system of globally coupled phase oscillators
have been experimentally realized using
electrochemical oscillators~\cite{ref:kiss02,ref:kiss07},
discrete chemical oscillators~\cite{ref:taylor09,ref:tinsley12}, and
mechanical oscillators~\cite{ref:pantaleone02,ref:martens13}.

To study the phase synchronization between macroscopic rhythms,
we recently formulated a theory for the collective phase description of macroscopic rhythms
emerging from coupled phase oscillators for the following three representative cases:
(A) phase coherent states in globally coupled noisy identical oscillators~\cite{ref:kawamura08,ref:ykawamura13a,ref:kawamura10a},
(B) partially phase-locked states in globally coupled noiseless nonidentical oscillators~\cite{ref:kawamura10b}, and
(C) fully phase-locked states in networks of coupled noiseless nonidentical oscillators~\cite{ref:kori09}.
The theory enables us to describe the dynamics of a macroscopic rhythm
by a single degree of freedom called the collective phase.
Accordingly,
different mathematical treatments were required for the physical situation in each case.
The keystone of the collective phase description method for each case is the following:
(A) the nonlinear Fokker-Planck equation~\cite{ref:kuramoto84},
(B) the Ott-Antonsen ansatz~\cite{ref:ott08,ref:ott09,ref:ott11}, and
(C) the Laplacian matrix~\cite{ref:boccaletti06,ref:arenas08,ref:dorogovtsev08,ref:barrat08}.
Here, we note that there exist several investigations
\cite{ref:masuda09a,ref:masuda09b,ref:masuda10,ref:ko09,ref:toenjes09,ref:cross12,ref:cross13}
related to case~(C).

In Ref.~\cite{ref:kawamura10a} for case~(A) and Ref.~\cite{ref:kawamura10b} for case~(B),
we investigated the phase synchronization between collective rhythms of globally coupled oscillator groups.
In particular,
the collective phase coupling function,
which determines the dynamics of the collective phase difference between the groups,
was systematically analyzed for sinusoidal coupling functions.
As a result, for both cases,
we found counter-intuitive phenomena
in which the groups can exhibit anti-phase collective synchronization
in spite of microscopic in-phase external coupling and vice versa.

\begin{figure*}
  \begin{center}
    \includegraphics[width=0.30\hsize,clip]{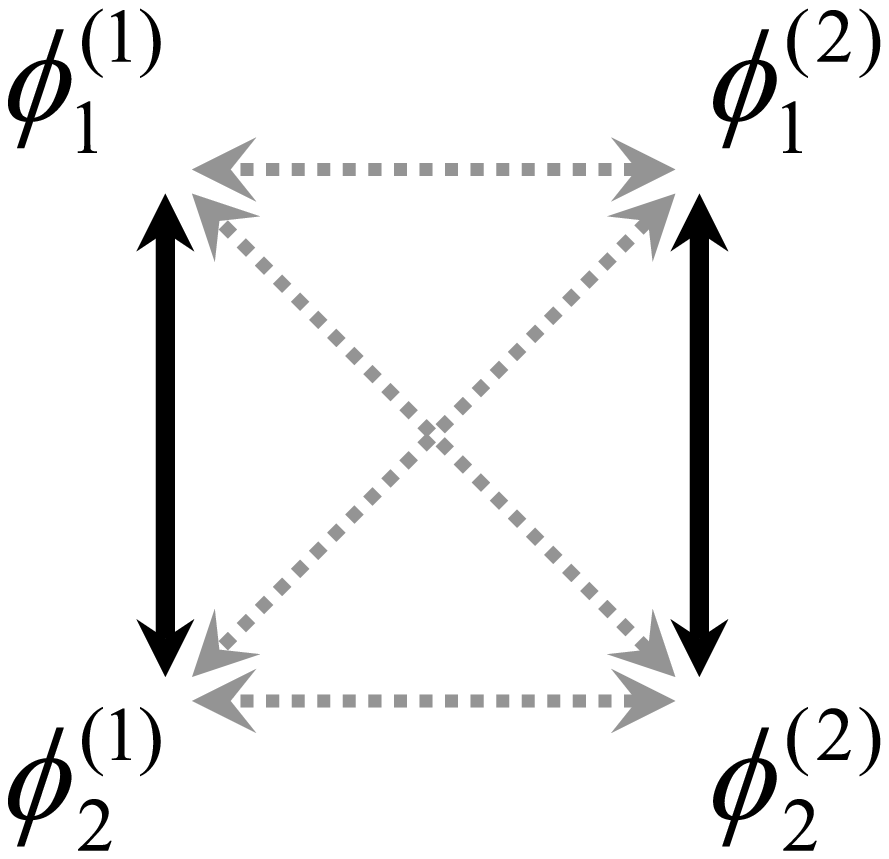}
    \caption{(Color online)
      Schematic diagram of two weakly interacting groups of two oscillators with global coupling.
      The microscopic internal and external couplings are represented by the solid and dotted arrows, respectively,
      whereas the self-coupling is not shown.
      The phase of the $j$-th oscillator in the $\sigma$-th group is denoted by $\phi_j^{(\sigma)}$.
    }
    \label{fig:1}
  \end{center}
\end{figure*}

In this paper,
we study the phase synchronization between collective rhythms of coupled oscillator groups
for case~(C).
We analytically derive the collective phase coupling function
for two weakly interacting groups of two oscillators with global sinusoidal coupling (see Fig.~\ref{fig:1}).
We thereby demonstrate counter-intuitive phenomena similar to those found in cases~(A) and (B),
that is, effective anti-phase (in-phase) collective synchronization
with microscopic in-phase (anti-phase) external coupling.
Therefore,
this paper and Refs.~\cite{ref:kawamura10a,ref:kawamura10b} are mutually complementary
and together provide a deeper understanding of the collective phase synchronization phenomena.

This paper is organized as follows.
In Sec.~\ref{sec:2},
we review the collective phase description of fully locked states.
In Sec.~\ref{sec:3},
we analyze weakly interacting groups of globally coupled two phase oscillators.
In Sec.~\ref{sec:4},
we perform further analytical calculations for the case of sinusoidal phase coupling.
In Sec.~\ref{sec:5},
we illustrate the collective phase coupling function for several representative cases.
In Sec.~\ref{sec:6},
we demonstrate collective phase synchronization by direct numerical simulations.
In App.~\ref{sec:A},
we also consider interacting groups of weakly coupled Stuart-Landau oscillators.
Concluding remarks are given in Sec.~\ref{sec:7}.

\section{Collective phase description of fully locked states} \label{sec:2}

In this section,
we review the collective phase description method
for fully phase-locked states in networks of coupled noiseless nonidentical oscillators
with an emphasis on the derivation of the collective phase coupling function.
More details and other applications of this collective phase description method
are given in Refs.~\cite{ref:kori09,ref:masuda09a,ref:masuda09b,ref:masuda10}.

We consider weakly interacting groups of coupled noiseless nonidentical phase oscillators
described by the following equation:
\begin{equation}
  \dot{\phi}_j^{(\sigma)}(t)
  = \omega_j
  + \sum_{k=1}^N \Gamma_{jk}\left( \phi_j^{(\sigma)} - \phi_k^{(\sigma)} \right)
  + \epsilon \sum_{k=1}^N \Gamma_{jk}^{\sigma\tau}\left( \phi_j^{(\sigma)} - \phi_k^{(\tau)} \right),
  \label{eq:model}
\end{equation}
for $j = 1, \cdots, N$ and $(\sigma, \tau) = (1, 2), (2, 1)$,
where $\phi_j^{(\sigma)}(t) \in \mathbb{S}^1$ is the phase of the $j$-th oscillator at time $t$
in the $\sigma$-th group consisting of $N$ oscillators
and $\omega_j$ is the natural frequency of the $j$-th phase oscillator.
The second term on the right-hand side represents the microscopic internal coupling within the same group,
while the third term represents the microscopic external coupling between the different groups.
The characteristic intensity of the external coupling is given by $\epsilon \geq 0$.
When the external coupling is absent, i.e., $\epsilon = 0$,
Eq.~(\ref{eq:model}) is assumed to have a stable fully phase-locked collective oscillation solution
\cite{ref:izhikevich07,ref:ermentrout10,ref:ermentrout92,ref:mirollo05}
\begin{equation}
  \phi_j^{(\sigma)}(t)
  = \Theta^{(\sigma)}(t) + \psi_j,
  \qquad
  \dot{\Theta}^{(\sigma)}(t)
  = \Omega
  = \omega_j + \sum_{k=1}^N \Gamma_{jk}\left( \psi_j - \psi_k \right),
  \label{eq:locked}
\end{equation}
where $\Theta^{(\sigma)}(t) \in \mathbb{S}^1$ is the collective phase at time $t$ for the $\sigma$-th group,
$\Omega$ is the collective frequency,
and the constants $\psi_j$ represent the relative phases of the individual oscillators for the fully phase-locked state.

When the external coupling is sufficiently weak, i.e., $\epsilon \ll 1$,
each group of oscillators obeying Eq.~(\ref{eq:model})
is always in the near vicinity of the fully phase-locked solution~(\ref{eq:locked}).
Therefore, we can approximately derive a collective phase equation in the following form~\cite{ref:kori09}:
\begin{equation}
  \dot{\Theta}^{(\sigma)}(t)
  = \Omega
  + \epsilon \digamma^{\sigma\tau}\left( \Theta^{(\sigma)} - \Theta^{(\tau)} \right),
  \label{eq:collective}
\end{equation}
where the {\it collective phase coupling function} is given by
\begin{equation}
  \digamma^{\sigma\tau}\left( \Theta^{(\sigma)} - \Theta^{(\tau)} \right)
  = \sum_{j=1}^N \sum_{k=1}^N U_j^\ast
  \Gamma_{jk}^{\sigma\tau}\left( \Theta^{(\sigma)} - \Theta^{(\tau)} + \psi_j - \psi_k \right).
  \label{eq:formula}
\end{equation}
Here, $U_j^\ast$ is the left zero eigenvector of the Jacobi matrix $L_{jk}$
at the fully phase-locked collective oscillation solution defined in Eq.~(\ref{eq:locked}).
The Jacobi matrix $L_{jk}$ is given by
\begin{equation}
  L_{jk}
  = \delta_{jk} \sum_{l \ne j} \Gamma_{jl}'\left( \psi_j - \psi_l \right)
  - \left( 1 - \delta_{jk} \right) \Gamma_{jk}'\left( \psi_j - \psi_k \right),
  \label{eq:Laplacian}
\end{equation}
which is a Laplacian matrix~\cite{ref:boccaletti06,ref:arenas08,ref:dorogovtsev08,ref:barrat08}.
That is, the Jacobi matrix $L_{jk}$ possesses the following property for each $j$:
$\sum_{k=1}^N L_{jk} = 0$.
In Eq.~(\ref{eq:Laplacian}),
we have used the Kronecker delta $\delta_{jk}$
and derivative notation $\Gamma_{jk}'(\phi) = d\Gamma_{jk}(\phi)/d\phi$.
We also note that
the Jacobi matrix $L_{jk}$ defined in Eq.~(\ref{eq:Laplacian}) is generally asymmetric and weighted.
Using the $(j,j)$-cofactor of the Jacobi matrix and the summation over the index $j$, i.e.,
\begin{equation}
  M_j = \det \hat{L}(j, j),
  \qquad
  M = \sum_{j=1}^N M_j,
  \label{eq:cofactor}
\end{equation}
the left zero eigenvector $U_j^\ast$ of the Jacobi matrix
that takes the form of the Laplacian matrix can be generally written
in the following form~\cite{ref:kori09,ref:masuda09a,ref:masuda09b,ref:masuda10}:
\begin{equation}
  U_j^\ast = \frac{M_j}{M},
  \qquad
  \sum_{j=1}^N U_j^\ast = 1.
  \label{eq:leftzero}
\end{equation}
In Eq.~(\ref{eq:cofactor}),
the matrix $\hat{L}(j, j)$ is the Jacobi matrix $\hat{L}$ with the $j$-th row and column removed,
and the cofactor $M_j$ is equal to the sum of the weights of all directed spanning trees rooted at the node $j$
according to the matrix tree theorem~\cite{ref:biggs97,ref:agaev00}.
Finally, we note that
the collective phase $\Theta^{(\sigma)}$ can be written in the following form~\cite{ref:kori09}:
\begin{equation}
  \Theta^{(\sigma)}
  = \sum_{j=1}^N U_j^\ast \left( \phi_j^{(\sigma)} - \psi_j \right),
  \label{eq:isochron}
\end{equation}
under the linear approximation of the isochron
\cite{ref:winfree80,ref:kuramoto84,ref:pikovsky01,ref:izhikevich07,ref:ermentrout10,ref:schultheiss12}.
In the following section,
we analyze globally-coupled two-oscillator systems
using this collective phase description method for fully locked states.

\section{Interacting groups of globally coupled two phase oscillators} \label{sec:3}

We first consider weakly interacting groups of globally coupled phase oscillators.
That is, the microscopic internal and external coupling functions are given by
\begin{align}
  \Gamma_{jk}\left( \phi_j^{(\sigma)} - \phi_k^{(\sigma)} \right)
  &= \frac{1}{N} \, \Gamma\left( \phi_j^{(\sigma)} - \phi_k^{(\sigma)} \right),
  \label{eq:global-in} \\
  \Gamma_{jk}^{\sigma\tau}\left( \phi_j^{(\sigma)} - \phi_k^{(\tau)} \right)
  &= \frac{1}{N} \, \Gamma^{\sigma\tau}\left( \phi_j^{(\sigma)} - \phi_k^{(\tau)} \right).
  \label{eq:global-ex}
\end{align}
In this global coupling case,
Eq.~(\ref{eq:model}) is written in the following form:
\begin{equation}
  \dot{\phi}_j^{(\sigma)}(t)
  = \omega_j
  + \frac{1}{N} \sum_{k=1}^N \Gamma\left( \phi_j^{(\sigma)} - \phi_k^{(\sigma)} \right)
  + \frac{\epsilon}{N} \sum_{k=1}^N \Gamma^{\sigma\tau}\left( \phi_j^{(\sigma)} - \phi_k^{(\tau)} \right).
  \label{eq:global}
\end{equation}

We further focus on the case in which the number of oscillators within each group is two, i.e., $N = 2$;
a schematic diagram of the case is shown in Fig.~\ref{fig:1}.
In this case, the internal dynamics for each group, i.e., Eq.~(\ref{eq:global}) with $\epsilon = 0$, is described as follows:
\begin{align}
  \dot{\phi}_1(t)
  &= \omega_1 + \frac{1}{2}\, \Gamma(0) + \frac{1}{2}\, \Gamma\left( \phi_1 - \phi_2 \right),
  \label{eq:phi1} \\
  \dot{\phi}_2(t)
  &= \omega_2 + \frac{1}{2}\, \Gamma(0) + \frac{1}{2}\, \Gamma\left( \phi_2 - \phi_1 \right),
  \label{eq:phi2}
\end{align}
where we dropped the group index $\sigma$ for simplicity.
From Eqs.~(\ref{eq:phi1}) and (\ref{eq:phi2}),
we obtain the following equation by subtraction:
\begin{equation}
  \frac{d}{dt} \Delta\phi(t)
  = \Delta\omega
  + \frac{1}{2}\, \Gamma\left(  \Delta\phi \right)
  - \frac{1}{2}\, \Gamma\left( -\Delta\phi \right),
  \label{eq:Dphi}
\end{equation}
where the phase difference $\Delta\phi(t)$ and frequency mismatch $\Delta\omega$ are defined as
\begin{equation}
  \Delta\phi(t) = \phi_1(t) - \phi_2(t),
  \qquad
  \Delta\omega = \omega_1 - \omega_2.
  \label{eq:Delta}
\end{equation}

Now, we assume that Eq.~(\ref{eq:Dphi}) has a fully phase-locked collective oscillation solution.
The phase difference of the stable phase-locked solution,
$\Delta\psi = \psi_1 - \psi_2$,
is determined by the following equation:
\begin{equation}
  \Delta\omega
  + \frac{1}{2}\, \Gamma\left(  \Delta\psi \right)
  - \frac{1}{2}\, \Gamma\left( -\Delta\psi \right)
  = 0.
  \label{eq:Dpsi}
\end{equation}
Using the phase difference $\Delta\psi$ obtained from Eq.~(\ref{eq:Dpsi}),
the collective frequency $\Omega$ is written in the following form:
\begin{equation}
  \Omega
  = \omega_1 + \frac{1}{2}\, \Gamma(0) + \frac{1}{2}\, \Gamma\left(  \Delta\psi \right)
  = \omega_2 + \frac{1}{2}\, \Gamma(0) + \frac{1}{2}\, \Gamma\left( -\Delta\psi \right).
  \label{eq:Omega}
\end{equation}
For these globally-coupled two-oscillator systems,
the Jacobi matrix $\hat{L}$ defined in Eq.~(\ref{eq:Laplacian}) is given by
\begin{equation}
  \hat{L} =
  \frac{1}{2}
  \begin{pmatrix}
     \Gamma'\left(  \Delta\psi \right)  &  -\Gamma'\left(  \Delta\psi \right) \\
    -\Gamma'\left( -\Delta\psi \right)  &   \Gamma'\left( -\Delta\psi \right) \\
  \end{pmatrix}.
  \label{eq:Laplacian2}
\end{equation}
Therefore, the cofactors of the Jacobi matrix are given by
\begin{equation}
  M_1 = \frac{\Gamma'\left( -\Delta\psi \right)}{2},
  \qquad
  M_2 = \frac{\Gamma'\left(  \Delta\psi \right)}{2},
  \label{eq:cofactor2}
\end{equation}
and the sum of these cofactors is written as
\begin{equation}
  M = M_1 + M_2
  = \frac{\Gamma'\left( -\Delta\psi \right) + \Gamma'\left( \Delta\psi \right)}{2}.
  \label{eq:summation}
\end{equation}
As found from Eq.~(\ref{eq:leftzero}),
using these cofactors and the sum,
Eq.~(\ref{eq:cofactor2}) and Eq.~(\ref{eq:summation}),
the left zero eigenvector $U_j^\ast$ is obtained as
\begin{equation}
  U_1^\ast = \frac{M_1}{M}
  = \frac{\Gamma'\left( -\Delta\psi \right)}{\Gamma'\left( -\Delta\psi \right) + \Gamma'\left( \Delta\psi \right)},
  \qquad
  U_2^\ast = \frac{M_2}{M}
  = \frac{\Gamma'\left(  \Delta\psi \right)}{\Gamma'\left( -\Delta\psi \right) + \Gamma'\left( \Delta\psi \right)}.
  \label{eq:leftzero2}
\end{equation}
Finally, we note that the Jacobi matrix $\hat{L}$ possesses
not only the zero eigenvalue but also the following non-zero eigenvalue:
\begin{equation}
  \lambda = M
  = \frac{\Gamma'\left( -\Delta\psi \right) + \Gamma'\left( \Delta\psi \right)}{2}.
  \label{eq:spectralgap}
\end{equation}
When the external coupling intensity is sufficiently small
compared to the absolute value of this non-zero eigenvalue,
i.e., $\epsilon \ll |\lambda|$,
the collective phase description is valid~\cite{ref:kori09}.

\section{Analytical formulas for the case of sinusoidal phase coupling} \label{sec:4}

In this section,
we consider the case of sinusoidal phase coupling functions for both microscopic internal and external couplings.
First, the microscopic internal phase coupling function is given by
\begin{equation}
  N \Gamma_{jk}\left( \phi \right)
  = \Gamma\left( \phi \right)
  = -\sin\left( \phi + \alpha \right),
  \qquad
  \left| \alpha \right| < \frac{\pi}{2},
  \label{eq:alpha}
\end{equation}
which is in-phase coupling (i.e., attractive).
By substituting Eq.~(\ref{eq:alpha}) into Eq.~(\ref{eq:Dpsi}),
the phase difference of the fully phase-locked state is obtained as
\begin{equation}
  \sin(\Delta\psi) = \eta,
  \qquad
  \eta \equiv \frac{\Delta\omega}{\cos\alpha},
  \qquad
  \left| \eta \right| < 1,
  \label{eq:solution}
\end{equation}
which indicates that the fully phase-locked solution emerge from a saddle-node bifurcation
and exists under the condition of $|\Delta\omega| < \cos\alpha$.
Owing to the in-phase coupling, i.e., Eq.~(\ref{eq:alpha}),
one solution of $|\Delta\psi| < \pi/2$ is stable,
and the other solution of $|\Delta\psi| > \pi/2$ is unstable.
Hereafter, the fully phase-locked solution indicates the stable one, $|\Delta\psi| < \pi/2$.
Substituting Eqs.~(\ref{eq:alpha}) and (\ref{eq:solution}) into Eq.~(\ref{eq:Omega}),
we obtain the collective frequency $\Omega$ as
\begin{equation}
  \Omega = \omega_1 - \frac{\eta}{2} \cos\alpha - \frac{1 + \sqrt{1 - \eta^2}}{2} \sin\alpha.
  \label{eq:frequency}
\end{equation}
Similarly,
substituting Eqs.~(\ref{eq:alpha}) and (\ref{eq:solution}) into Eq.~(\ref{eq:cofactor2}),
we obtain the cofactors as follows:
\begin{equation}
  M_1 = \frac{- \sqrt{1 - \eta^2}\, \cos\alpha - \eta \sin\alpha}{2},
  \qquad
  M_2 = \frac{- \sqrt{1 - \eta^2}\, \cos\alpha + \eta \sin\alpha}{2},
  \label{eq:determinant}
\end{equation}
which yield $M = - \sqrt{1 - \eta^2} \, \cos\alpha$.
From Eqs.~(\ref{eq:leftzero2}) and (\ref{eq:determinant}),
the left zero eigenvector $U_j^\ast$ is thus written as
\begin{equation}
  U_1^\ast = \frac{1}{2} \left( 1 + \frac{\eta\tan\alpha}{\sqrt{1 - \eta^2}} \right),
  \qquad
  U_2^\ast = \frac{1}{2} \left( 1 - \frac{\eta\tan\alpha}{\sqrt{1 - \eta^2}} \right).
  \label{eq:eigenvector}
\end{equation}
In addition,
the non-zero eigenvalue $\lambda$ defined in Eq.~(\ref{eq:spectralgap}) is obtained as
\begin{equation}
  \lambda = - \sqrt{1 - \eta^2}\, \cos\alpha.
  \label{eq:lambda}
\end{equation}

Next, the microscopic external phase coupling function is given by
\begin{equation}
  N \Gamma_{jk}^{\sigma\tau}\left( \phi \right)
  = \Gamma^{\sigma\tau}\left( \phi \right)
  = -\sin\left( \phi + \beta \right),
  \label{eq:beta}
\end{equation}
which can be either in-phase coupling (i.e., attractive) under the condition of $|\beta| < \pi/2$
or anti-phase coupling (i.e., repulsive) under the condition of $|\beta| > \pi/2$.
By plugging Eqs.~(\ref{eq:solution}), (\ref{eq:eigenvector}), and (\ref{eq:beta}) into Eq.~(\ref{eq:formula}),
the collective phase coupling function takes the following form:
\begin{equation}
  \digamma^{\sigma\tau}\left( \Theta \right) = -\rho\sin\left( \Theta + \delta \right),
  \label{eq:digamma}
\end{equation}
where the complex number with modulus $\rho$ and argument $\delta$ is given by
\begin{align}
  \rho e^{i\delta}
  =&\, \frac{1}{2} \left[ \left( 1 + \sqrt{1 - \eta^2} \right) \cos\beta
    - \frac{\eta^2}{\sqrt{1 - \eta^2}} \tan\alpha \sin\beta \right]
  \nonumber \\
  &+ \frac{i}{2} \left[ \left( 1 + \sqrt{1 - \eta^2} \right) \sin\beta
    + \frac{\eta^2}{\sqrt{1 - \eta^2}} \tan\alpha \cos\beta \right].
  \label{eq:type}
\end{align}
This formula is the main result of the present paper.
It determines the collective phase coupling function
for two weakly interacting groups of two oscillators with global sinusoidal coupling.
The coupling type can be found from the real part, i.e.,
\begin{equation}
  \rho \cos \delta
  = \frac{1}{2} \left[ \left( 1 + \sqrt{1 - \eta^2} \right) \cos\beta
    - \frac{\eta^2}{\sqrt{1 - \eta^2}} \tan\alpha \sin\beta \right],
  \label{eq:type2}
\end{equation}
where $\rho \cos \delta > 0$ and $\rho \cos \delta < 0$
indicate in-phase and anti-phase couplings, respectively.
Finally, we note that Eq.~(\ref{eq:type2}) possesses origin symmetry in the $\alpha$-$\beta$ plane.

\section{Type of the collective phase coupling function for representative cases} \label{sec:5}

In this section,
we study the type of the collective phase coupling function for the following five representative cases.

(i) The first case is $\eta = 0$,
which indicates that two oscillators within each group are identical, i.e., $\Delta\omega = 0$.
Substituting $\eta = 0$ into Eq.~(\ref{eq:type}),
we obtain the following result:
\begin{equation}
  \eta = 0,
  \qquad
  \rho e^{i\delta} = e^{i\beta}.
  \label{eq:case1}
\end{equation}
That is, the collective phase coupling function
is the same as the microscopic external phase coupling function, i.e.,
$\digamma^{\sigma\tau}(\Theta) = \Gamma^{\sigma\tau}(\Theta) = -\sin(\Theta + \beta)$.

(ii) The second case is $|\eta| \simeq 1$,
which indicates the proximity of the saddle-node bifurcation point,
i.e., the onset of fully phase-locked collective oscillation.
Substituting $|\eta| \simeq 1$ into Eq.~(\ref{eq:type}),
we obtain the following result:
\begin{equation}
  \left| \eta \right| \simeq 1,
  \qquad
  \rho e^{i\delta}
  \simeq -\frac{1}{2} \left[ \frac{\eta^2}{\sqrt{1 - \eta^2}} \tan\alpha \sin\beta \right]
  + \frac{i}{2} \left[ \frac{\eta^2}{\sqrt{1 - \eta^2}} \tan\alpha \cos\beta \right].
  \label{eq:case2}
\end{equation}
For the case of $|\eta| \to 1$ (excluding $\alpha = 0$),
the amplitude of the collective phase coupling becomes infinity, i.e., $\rho \to \infty$.
Here, we note that this property for the fully phase-locked states is quite different from
those for phase coherent states and partially phase-locked states~\cite{ref:kawamura10a,ref:kawamura10b}.
For the latter two states,
the amplitude of the collective phase coupling is finite at the onset of collective oscillations.
This difference in the properties results from the difference of bifurcations.
The fully phase-locked states emerge from saddle-node bifurcations as mentioned above,
whereas the phase coherent states and partially phase-locked states
emerge from supercritical Hopf bifurcations~\cite{ref:kawamura10a,ref:kawamura10b}.

(iii) The third case is $\alpha = 0$,
which yields a microscopic antisymmetric internal coupling function.
For this case, $\eta = \Delta\omega$.
Substituting $\alpha = 0$ into Eq.~(\ref{eq:type}),
we obtain the following result:
\begin{equation}
  \alpha = 0,
  \qquad
  \rho e^{i\delta} = \frac{1 + \sqrt{1 - \eta^2}}{2} e^{i\beta}.
  \label{eq:case3}
\end{equation}
That is, the phase shift $\delta$ of the collective phase coupling function
is the same as the phase shift $\beta$ of the microscopic external phase coupling function.

(iv) The fourth cases are special values of $\beta$.
Substituting $\beta = 0$, $\pm\pi$, $\pm\pi/2$ into Eq.~(\ref{eq:type}),
we obtain the following results:
\begin{align}
  \beta = 0,
  \qquad
  \rho e^{i\delta}
  &= +\frac{1}{2} \left[ 1 + \sqrt{1 - \eta^2} \right]
  + \frac{i}{2} \left[ \frac{\eta^2}{\sqrt{1 - \eta^2}} \tan\alpha \right],
  \label{eq:case4a} \\
  \beta = \pm\pi,
  \qquad
  \rho e^{i\delta}
  &= -\frac{1}{2} \left[ 1 + \sqrt{1 - \eta^2} \right]
  - \frac{i}{2} \left[ \frac{\eta^2}{\sqrt{1 - \eta^2}} \tan\alpha \right],
  \label{eq:case4b} \\
  \beta = \pm\frac{\pi}{2},
  \qquad
  \rho e^{i\delta}
  &= \mp \frac{1}{2} \left[ \frac{\eta^2}{\sqrt{1 - \eta^2}} \tan\alpha \right]
  \pm \frac{i}{2} \left[ 1 + \sqrt{1 - \eta^2} \right].
  \label{eq:case4c}
\end{align}
For microscopic antisymmetric external coupling functions, i.e., $\beta = 0$, $\pm\pi$,
the type of the collective phase coupling function
coincides with that of the microscopic external coupling function.
In contrast, for microscopic symmetric external coupling functions, i.e., $\beta = \pm\pi/2$,
the type of the collective phase coupling function
is determined by the sign of the microscopic internal coupling parameter $\alpha$.

(v) The fifth case is $\beta = \alpha$,
which indicates that the microscopic external coupling has the same phase shift as the microscopic internal one.
Substituting $\beta = \alpha$ into Eq.~(\ref{eq:type}),
we obtain the following result:
\begin{equation}
  \beta = \alpha,
  \qquad
  \rho e^{i\delta}
  = \frac{1}{2} \left[ \frac{1 + \sqrt{1 - \eta^2}}{\sqrt{1 - \eta^2}} \cos\alpha
    - \frac{\eta^2}{\sqrt{1 - \eta^2}} \frac{1}{\cos\alpha} \right]
  + \frac{i}{2} \left[ \frac{1 + \sqrt{1 - \eta^2}}{\sqrt{1 - \eta^2}} \sin\alpha \right].
  \label{eq:case5}
\end{equation}
From the condition of $|\alpha| < \pi / 2$,
both microscopic internal and external coupling functions are in-phase coupling.
However, the type of the collective phase coupling function is anti-phase coupling
under the following condition:
\begin{equation}
  \cos^2 \alpha < \frac{\eta^2}{1 + \sqrt{1 - \eta^2}}.
  \label{eq:case5eta}
\end{equation}
For the case of $|\eta| \to 1$,
the above condition becomes $\cos^2 \alpha < 1$,
which is satisfied for all $\alpha$ except for $\alpha = 0$.

\section{Collective phase synchronization between two interacting groups} \label{sec:6}

\begin{figure*}
  \begin{center}
    \includegraphics[width=0.55\hsize,clip]{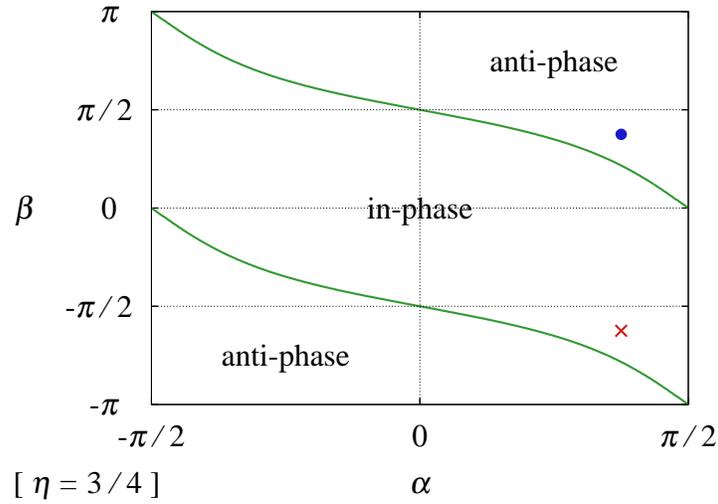}
    \caption{(Color online)
      Effective type of phase coupling between collective rhythms
      of fully locked oscillator groups with
      $\alpha \in ( -\pi / 2, \pi / 2 )$,
      $\beta \in [ -\pi, \pi ]$, and
      $\eta = 3 / 4$.
      The solid curves are analytically determined by Eq.~(\ref{eq:type2}),
      i.e., $\rho \cos \delta = 0$.
      The filled circle ($\bullet$) indicates
      $\alpha = \beta = 3 \pi / 8$
      corresponding to Fig.~\ref{fig:3}(a) and Fig.~\ref{fig:4}.
      The times sign ($\times$) indicates
      $\alpha = 3 \pi / 8$ and $\beta = -5 \pi / 8$
      corresponding to Fig.~\ref{fig:3}(b).
    }
    \label{fig:2}
  \end{center}
\end{figure*}

Now, we study counter-intuitive cases under the condition of $\eta = 3/4$.
The type of the collective phase coupling function is shown in Fig.~\ref{fig:2},
where the solid curves are determined by Eq.~(\ref{eq:type2}), i.e., $\rho \cos \delta = 0$.
Here, we note that the type of the collective phase coupling function
can be different from that of the microscopic external phase coupling function.
Two sets of parameters, which were used in Fig.~\ref{fig:3}, are also shown in Fig.~\ref{fig:2}.

Two groups of two-oscillators exhibiting phase-locked states
were separately prepared with their corresponding phases being nearly identical.
Then, these states were used as the initial condition in Fig.~\ref{fig:3}(a).
In spite of the microscopic in-phase external coupling, $\beta = 3 \pi / 8$,
the external phase difference $| \phi_1^{(1)} - \phi_1^{(2)} |$ approached $\pi$ after some time;
this indicates anti-phase collective synchronization between the groups.
In contrast,
Fig.~\ref{fig:3}(b) shows in-phase collective synchronization between the groups
in spite of the microscopic anti-phase external coupling, $\beta = -5 \pi / 8$.

Finally, it should be noted that
we can also consider interacting groups of weakly coupled Stuart-Landau oscillators
as mentioned in App.~\ref{sec:A}.

\section{Concluding remarks} \label{sec:7}

In this paper,
we considered the phase synchronization between collective rhythms of fully locked oscillator groups,
clarified the relation between the collective phase coupling and microscopic external phase coupling functions,
analytically determined the type of the collective phase coupling function
for weakly interacting groups of two oscillators with global sinusoidal coupling,
and demonstrated that the groups can exhibit anti-phase (in-phase) collective synchronization
in spite of microscopic in-phase (anti-phase) external coupling.
The theoretical predictions were successfully confirmed by direct numerical simulations
of the phase oscillator model and Stuart-Landau oscillator model.

In Refs.~\cite{ref:kawamura10a,ref:kawamura10b},
we investigated the phase synchronization between collective rhythms
of globally coupled oscillator groups under two typical situations:
phase coherent states in the noisy identical case~\cite{ref:kawamura10a}
and
partially phase-locked states in the noiseless nonidentical case~\cite{ref:kawamura10b}.
In particular, we found the counter-intuitive phenomena similar to the results in this paper.
That is, weakly interacting groups can exhibit anti-phase collective synchronization
in spite of microscopic in-phase external coupling and vice versa.
Here, we note that
these three papers considered different physical situations and utilized different mathematical methods,
but arrived at the similar counter-intuitive phenomena.

We also remark that
fully phase-locked states emerge from a finite number of oscillators~\cite{ref:izhikevich07,ref:ermentrout10};
even two is possible as actually studied in this paper.
In contrast,
phase coherent states and partially phase-locked states
emerge from a large population of oscillators~\cite{ref:kuramoto84};
the number of oscillators is infinite in theory.
From this point of view,
fully phase-locked states can be more easily realized in experiments such as
electrochemical oscillators~\cite{ref:kiss02,ref:kiss07},
discrete chemical oscillators~\cite{ref:taylor09,ref:tinsley12}, and
mechanical oscillators~\cite{ref:pantaleone02,ref:martens13}.
We hope that the counter-intuitive phenomena studied in this paper,
i.e., effective anti-phase (in-phase) collective synchronization
with microscopic in-phase (anti-phase) external coupling,
will be experimentally confirmed in the near future
and that the formula~(\ref{eq:type}) will help in such experiments.

\begin{acknowledgments}
  The author is grateful to Yoshiki Kuramoto, Hiroya Nakao, Hiroshi Kori, and Kensuke Arai for fruitful discussions.
  This work was supported by JSPS KAKENHI Grant Number 25800222.
\end{acknowledgments}

\appendix

\section{Interacting groups of weakly coupled Stuart-Landau oscillators} \label{sec:A}

\begin{figure*}
  \begin{center}
    \includegraphics[width=1.00\hsize,clip]{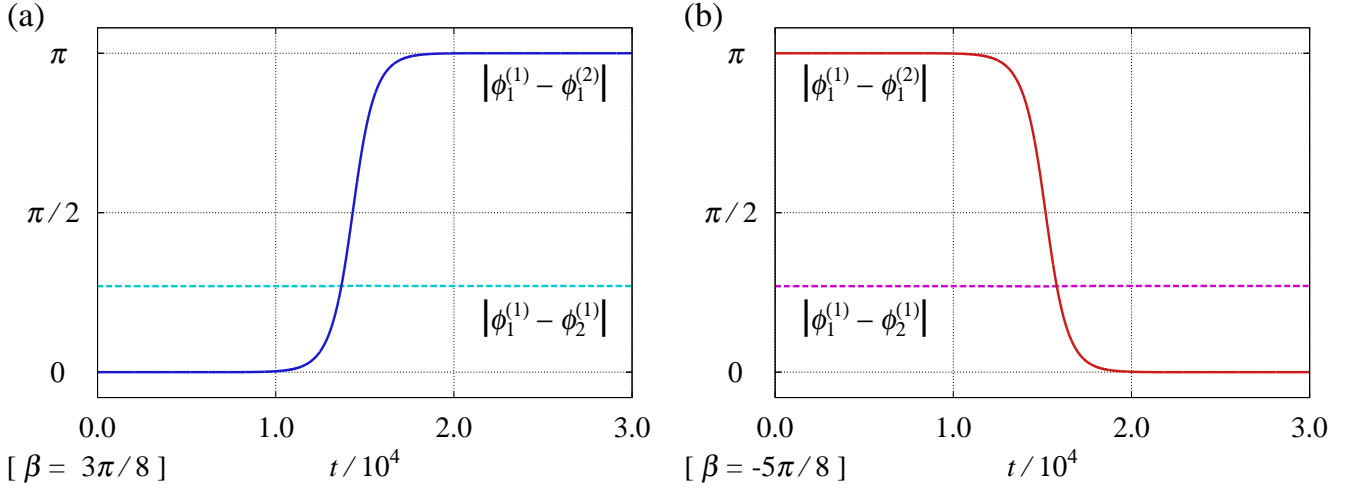}
    \caption{(Color online)
      Interacting groups of phase oscillators.
      Time evolution of the internal and external phase differences, i.e.,
      $| \phi_1^{(1)} - \phi_2^{(1)} |$ and $| \phi_1^{(1)} - \phi_1^{(2)} |$.
      The other internal and external phase differences are approximated as
      $| \phi_1^{(2)} - \phi_2^{(2)} | \simeq | \phi_1^{(1)} - \phi_2^{(1)} |$
      and
      $| \phi_2^{(1)} - \phi_2^{(2)} | \simeq | \phi_1^{(1)} - \phi_1^{(2)} |$,
      respectively.
      The collective phase difference is approximated as the external phase difference, i.e.,
      $| \Theta^{(1)} - \Theta^{(2)} | \simeq | \phi_1^{(1)} - \phi_1^{(2)} |$.
      The parameters are
      $\alpha = 3 \pi / 8$,
      $\omega_1 = 3 \cos(\alpha) / 4$,
      $\omega_2 = 0$, and
      $\epsilon = 0.001$.
      (a) Effective anti-phase collective synchronization
      with microscopic in-phase external coupling, $\beta = 3 \pi / 8$.
      (b) Effective in-phase collective synchronization
      with microscopic anti-phase external coupling, $\beta = -5 \pi / 8$.
    }
    \label{fig:3}
  \end{center}
\end{figure*}

\begin{figure*}
  \begin{center}
    \includegraphics[width=1.00\hsize,clip]{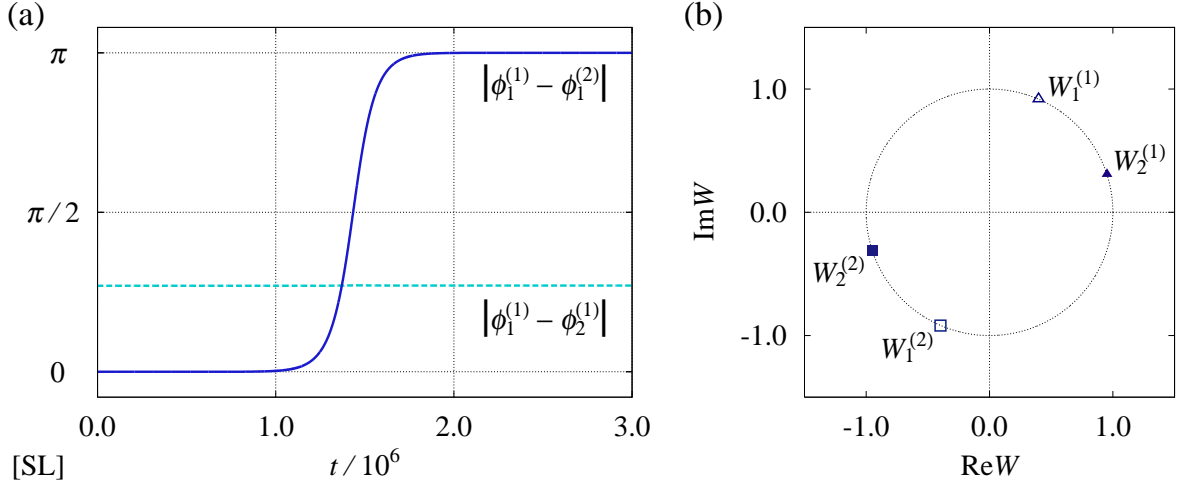}
    \caption{(Color online)
      Interacting groups of weakly coupled Stuart-Landau (SL) oscillators.
      Effective anti-phase collective synchronization
      with microscopic in-phase external coupling.
      The parameters are
      $K = J = r \cos(a)$,
      $c_1 = c_3 = 0$,
      $c_2 = \tan(a)$,
      $b_1 = c_2 + 3 r \cos(a) / 4$,
      $b_2 = c_2$,
      and
      $\epsilon = 0.001$,
      where
      $r = 0.01$
      and
      $a = 3 \pi / 8$.
      (a) Time evolution of the internal and external phase differences, i.e.,
      $| \phi_1^{(1)} - \phi_2^{(1)} |$ and $| \phi_1^{(1)} - \phi_1^{(2)} |$.
      (b) Snapshot of the asymptotic state of individual oscillators, i.e.,
      $W_1^{(1)}$, $W_2^{(1)}$, $W_1^{(2)}$, and $W_2^{(2)}$.
    }
    \label{fig:4}
  \end{center}
\end{figure*}

In this appendix,
we consider interacting groups of globally coupled Stuart-Landau oscillators described by the following equation:
\begin{equation}
  \dot{W}_j^{(\sigma)}(t)
  = (1 + i b_j) W_j^{(\sigma)} - (1 + i c_2) \left| W_j^{(\sigma)} \right|^2 W_j^{(\sigma)}
  + \frac{K (1 + i c_1)}{N} \sum_{k=1}^N W_k^{(\sigma)}
  + \frac{\epsilon J (1 + i c_3)}{N} \sum_{k=1}^N W_k^{(\tau)},
  \label{eq:A:model}
\end{equation}
for $j = 1, \cdots, N$ and $(\sigma, \tau) = (1, 2), (2, 1)$,
where $W_j^{(\sigma)}(t) \in \mathbb{C}^1$ is the complex amplitude of the $j$-th limit-cycle oscillator at time $t$
in the $\sigma$-th group consisting of $N$ oscillators.
The first and second terms on the right-hand side represent the intrinsic dynamics of each oscillator,
the third term represents the microscopic internal coupling within the same group,
and the fourth term represents the microscopic external coupling between the different groups.
When the internal and external couplings are sufficiently weak
compared to the absolute value of the amplitude Floquet exponent,
we can approximately derive a phase equation in the following form~\cite{ref:kuramoto84}:
\begin{equation}
  \dot{\phi}_j^{(\sigma)}(t)
  = \omega_j
  - \frac{P_K}{N} \sum_{k=1}^N \sin\left( \phi_j^{(\sigma)} - \phi_k^{(\sigma)} + \alpha \right)
  - \frac{\epsilon P_J}{N} \sum_{k=1}^N \sin\left( \phi_j^{(\sigma)} - \phi_k^{(\tau)} + \beta \right),
  \label{eq:A:phase}
\end{equation}
where the parameters of phase oscillators are given by
\begin{align}
  \omega_j
  &= b_j - c_2,
  \label{eq:A:omega} \\
  P_K\, e^{i \alpha}
  &= K (1 + i c_2) (1 - i c_1),
  \label{eq:A:alpha} \\
  P_J\, e^{i \beta}
  &= J (1 + i c_2) (1 - i c_3).
  \label{eq:A:beta}
\end{align}
The phase of each Stuart-Landau oscillator is given by the following equation
\cite{ref:winfree80,ref:kuramoto84,ref:pikovsky01,ref:izhikevich07,ref:ermentrout10,ref:schultheiss12}:
$\phi = \arg W - c_2 \ln |W|$.
As in the main text,
we focus on the case in which the number of oscillators within each group is two, i.e., $N = 2$.
Using the following constants, $r = 0.01$ and $a = 3 \pi / 8$,
the parameters of the Stuart-Landau oscillators are fixed at
$K = J = r \cos(a)$,
$c_1 = c_3 = 0$,
$c_2 = \tan(a)$,
$b_1 = c_2 + 3 r \cos(a) / 4$,
and
$b_2 = c_2$.
Under these conditions,
the parameters of the phase oscillators are obtained as
$P_K = P_J = r = 0.01$,
$\alpha = \beta = a = 3 \pi / 8$,
$\omega_1 = 3 r \cos(a) / 4$,
and
$\omega_2 = 0$,
which correspond to the parameters in Fig.~\ref{fig:3}(a).
In particular, we note that
$\eta = (\Delta\omega) / (P_K \cos\alpha) = (3 r \cos(a) / 4) / (r \cos(a)) = 3 / 4$.
The external coupling intensity is fixed at $\epsilon = 0.001$.
The direct numerical simulation result of Eq.~(\ref{eq:A:model}) is shown in Fig.~\ref{fig:4}.
Similarly to Fig.~\ref{fig:3}(a),
Fig.~\ref{fig:4} shows anti-phase collective synchronization between the groups
in spite of the microscopic in-phase external coupling.

\clearpage



\clearpage

\clearpage

\end{document}